\documentclass{aa}  

\usepackage{graphicx}
\usepackage{txfonts}
\usepackage{multirow}

\begin{document}

   \title{Exploring galaxy evolution with generative models}


   \author{Kevin Schawinski
          \inst{1}
          \and
          M. Dennis Turp\inst{1}\fnmsep
          \and
          Ce Zhang\inst{2}
          }

   \institute{Institute for Particle Physics and Astrophysics, Department of Physics, ETH Zurich, Wolfgang-Pauli-Strasse 27, CH-8093, Z\"{u}rich, Switzerland\\
              \email{kevin.schawinski@phys.ethz.ch, dturp@student.ethz.ch}
         \and
             Systems Group, Department of Computer Science, ETH Zurich, Universit\"{a}tstrasse 6, CH-8006, Z\"{u}rich, Switzerland\\
             \email{ce.zhang@inf.ethz.ch}
             }

   \date{Received July 9, 2018; accepted August 6, 2018}

 
  \abstract
   {Generative models open up the possibility to interrogate scientific data in a more data-driven way.}
   {We propose a method that uses generative models to explore hypotheses in astrophysics and other areas. We use a neural network to show how we can independently manipulate physical attributes by encoding objects in latent space. }
   {By learning a latent space representation of the data, we can use this network to forward model and explore hypotheses in a data-driven way. We train a neural network to generate artificial data to test hypotheses for the underlying physical processes. }
   {We demonstrate this process using a well-studied process in astrophysics, the quenching of star formation in galaxies as they move from low- to high-density environments. This approach can help explore astrophysical and other phenomena in a way that is different from current methods based on simulations and observations. }
   {}


   \maketitle
%

\section{Introduction}
Many objects of interest in astrophysics appear effectively static as the relevant characteristic timescales are far beyond human lifetimes. For this reason, there are generally two approaches researchers take to understand the formation and evolution of objects such as galaxies and quasars: they either take observations and fit models to the data, or they propose some underlying physical model and implement it in a simulation. Observations are limited by the underlying processes we can infer from them, and simulations make assumptions for the processes modeled and are often computationally very expensive.

\begin{figure}
\centering
\includegraphics[width=1.0\linewidth]{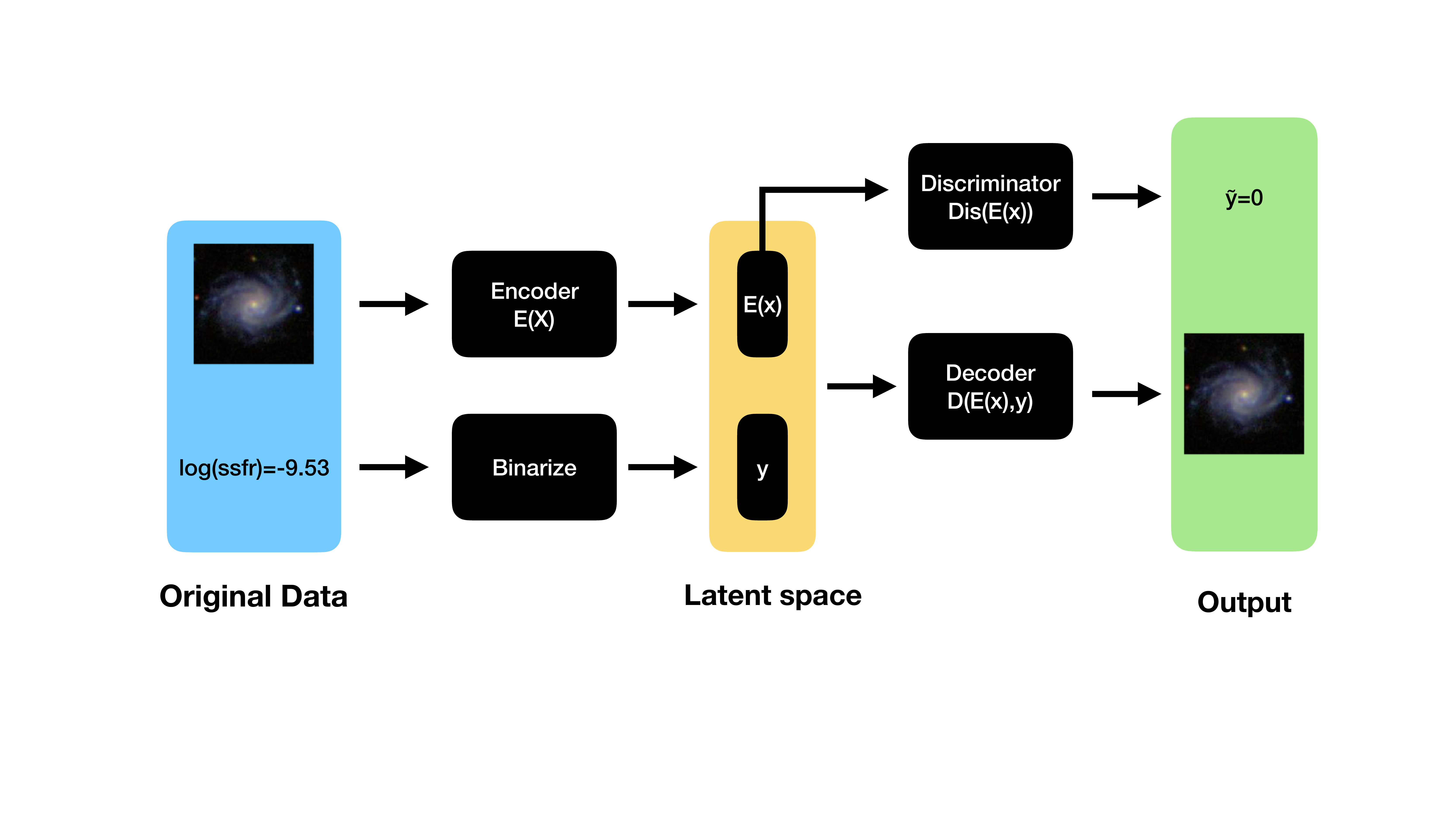}
\includegraphics[width=1.0\linewidth]{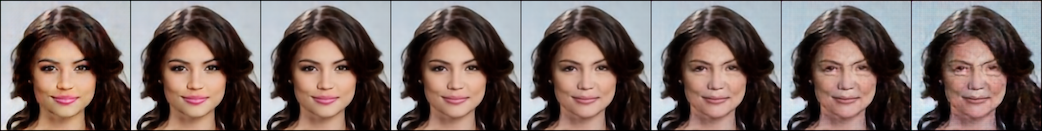}
\includegraphics[width=1.0\linewidth]{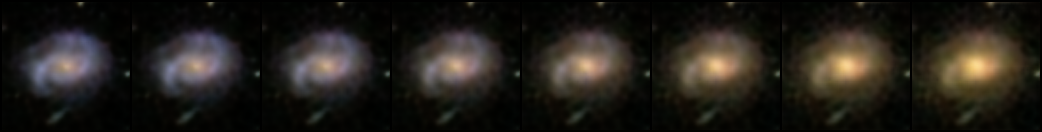}
\caption{Fader network architecture: Original galaxy images are input to an encoder $E(x)$ which performs a mapping to a latent space of fixed dimension. The associated physical property is binarized into a label $y$. The parametres $E(x)$ and y are input to a decoder $D(E(x),y)$ which tries to reconstruct the original input image. The discriminator $Dis(E(x))$ tries to predict the label $y$ from the latent code $E(x)$. Below, we show two examples of changing a single attribute in latent space using a fader network: the aging of a human face learned from age labels (using a pretrained model; \citealt{2017arXiv170600409L}) , and the lowering of the sSFR of a galaxy using sSFR labels. }
\label{fig:network}
\end{figure}

On the other hand,  an emerging technique \citep{2017arXiv170600409L, EURECOM+5134, NIPS2016_6051, DBLP:conf/conll/BowmanVVDJB16, kingma2013auto, rezende2014stochastic} has recently been developed by the machine-learning community that can perform tasks such as taking a  photo of a human being as input, and manipulating a certain attribute (e.g. age) and generating a new photo of the same person (See Figure \ref{fig:network}). In this paper, we ask the following question: {\em Can this method be used as a way to explore scientific hypotheses in a purely data-driven way? Can we achieve a similar physical understanding by using such a method?}

\begin{figure*}
\centering
\includegraphics[width=0.7\linewidth]{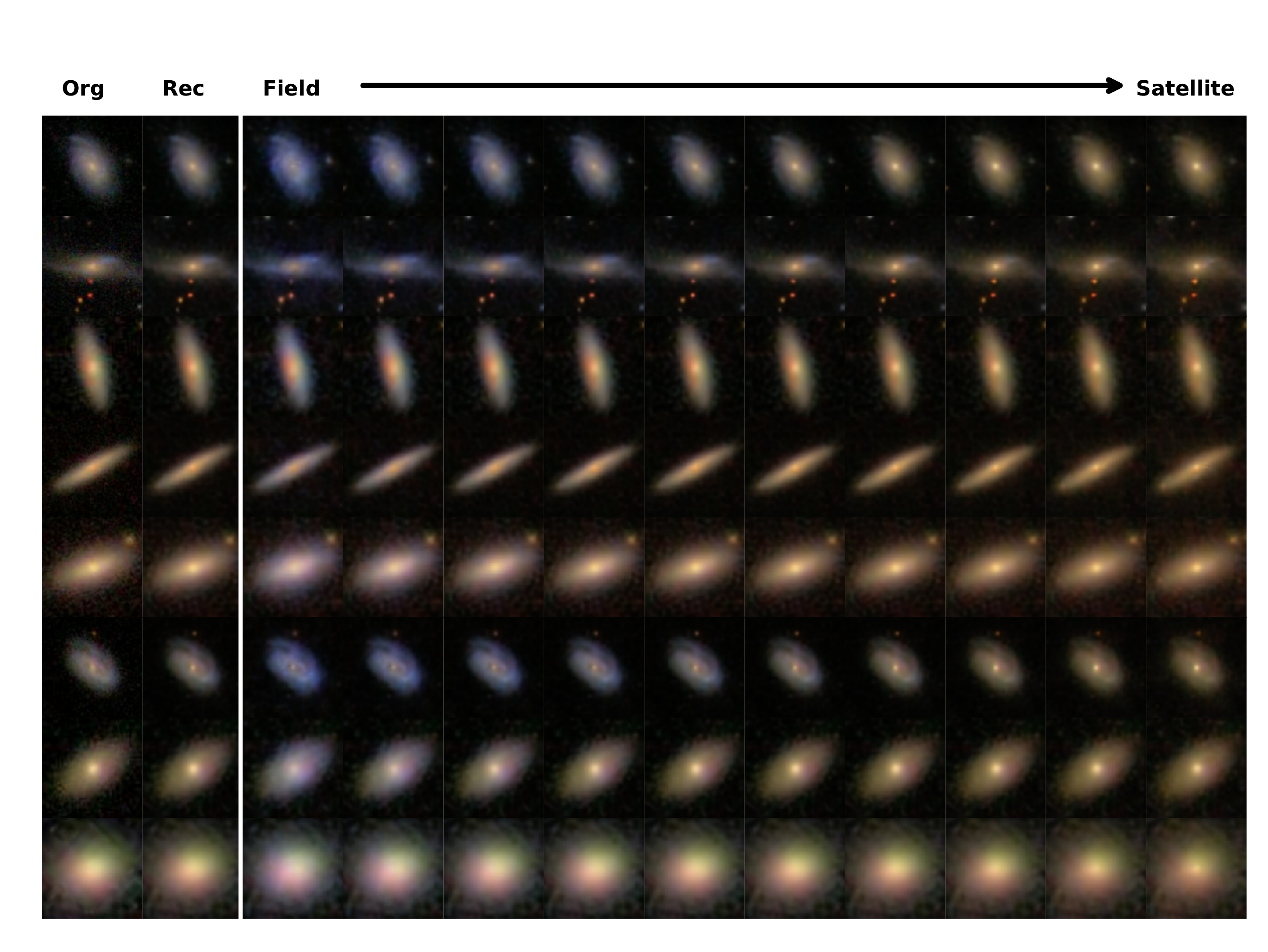}
\caption{Hypothesis of taking field galaxies and turning them into satellites. The \textit{Org} column shows the original galaxy image, and the \textit{Rec} column shows the best reconstruction of the original image. The remainder of each row shows the transformation from field galaxy to satellite as predicted by our environment Fader network. This figure represents our hypothesis-generation step as we see that when galaxies become satellites, they become redder, and their structure changes to become more centrally concentrated.}
\label{fig:hypothesis}
\end{figure*}

What we propose here is that we can take the data and interrogate the network trained on the data to explore plausible hypotheses. We start with a neural network trained on a set of objects associated with a given measured physical property. Once trained, we can then use the network in two ways: we can \textit{encode} a real object to a latent space, that is, obtain its representation in latent space by its latent space vector. We can also go the other way and construct a latent space vector and have the network \textit{decode} the corresponding object into real space. This latent space contains a model of all the salient features of the objects the network is trained on, and so for the network to perform this transformation well, it needs to learn the \textit{most salient features} of a group of objects. By changing the latent space vector entries, we can \textit{walk} in latent space and so transform objects from one state to another (Figure \ref{fig:network})\citep{2017arXiv170600409L, EURECOM+5134, NIPS2016_6051, DBLP:conf/conll/BowmanVVDJB16, kingma2013auto, rezende2014stochastic}. 

Using this structure, we can isolate the action of a parameter and observe its effect on the data by varying it independently of other properties. Suppose we start with a population of objects $A$ which we suspect evolves into population $B$. Now we can ask two questions: 
\begin{enumerate}
\item What are the changes that we observe when an object $a\in A$ evolves into an object $b\in B$?
\item What are the physical parameters that can explain these changes?
\end{enumerate}
Using our proposed method we can address both questions in a data-driven way:
we first train a network on populations $A$ and $B$ to visualize the differences between those two data distributions, {that is,} we can transform individual objects $a\in A \Rightarrow T(a)\in B $. Comparing $a$ and $T(a)$ we can use our domain knowledge and hypothesize possible physical parameters that can explain the changes that we see. With our set of possible parameters, $x,y,z,$ we can train the network to perform transformations $X,Y,Z$ based on these parameters. We then apply these transforms and compare $X(A),Y(A),Z(A)$ to $B$ and use some statistical measure to tell us how different they are from $B$.  If, for example, transform $X(A)$ is the closest to $B$ it is most likely to be a good explanation for, though not a proof of, how A evolves into B. If none of the transforms are sufficient, for example, the distance between the data distributions is too large, then either all transforms are not a good explanation, or alternatively the training data or network structure was not sufficient to learn $X,Y,Z$ well enough to test it, a possibility which affects all simulation-based hypotheses. The ability to generate and test hypotheses using this approach is limited by the available labels; a process could depend on parameters traced by multiple labels, or by parameters not captured by any of the available labels.

\section{Method} 
We use the Fader network \citep{2017arXiv170600409L} architecture to demonstrate how we can use this approach to test hypotheses in real astrophysical settings. The Fader network (Figure \ref{fig:network}) is based on an encoder-decoder structure with a domain adversarial aspect that allows us to learn and manipulate images based on physical properties which have to be converted into binary labels. The key is that the Fader network is able to learn and visualise differences between two data distributions. Images $x$ are input to a neural network called encoder $E(x)$. The encoder $E(x)$ performs a mapping from the space of images to a latent representation of fixed dimension. The decoder $D(E(x),y)$, also a neural network, tries to reconstruct the image $x$ given its latent representation $E(x)$ and its label $y$ (e.g. binary sSFR, dust). Thus the latent space is split into two parts. One part which should contain all the information on the labels and another part which should contain all the salient information needed to reconstruct the object. To perform this disentanglement, another neural network called discriminator $Dis(E(x))$ is trained to predict the label $y$ from the latent code $E(x)$. Given pairs of images and binary labels $\{x,y\}$ the Fader network tries to minimise two objectives:
\begin{equation} \label{eq1}
\mathcal{L}_{ae}=-\frac{1}{m}\sum \left\| D(E(x),y) - x \right\|_2^2 - \lambda_{E}\log(P(1-y|E(x)))
,\end{equation}
\begin{equation} \label{eq2}
\mathcal{L}_{dis}=-\frac{1}{m}\sum\log(P(y|E(x)))
.\end{equation}
The first term in $\mathcal{L}_{ae}$ is the reconstruction loss and measures how well the auto-encoder can reconstruct the original input. The second term is the domain adversarial component and by minimising it should become impossible to predict the physical property/label $y$ from the latent code $E(x)$.  At the same time the discriminator becomes better at predicting the physical property/label from the latent code $E(x)$ by minimizing $\mathcal{L}_{dis}$. This adversarial interplay between the two loss functions is what allows the fader network to disentangle salient and label information in the decoding process. During inference, the property/label information can be continuous and changing it will resemble what the original image would look like with the changed physical property/label.

\section{Specific scientific application}  
As a demonstration, we choose the question of satellite quenching in galaxy formation. This problem has several advantages: it is relatively well understood, it relies on changes in the imaging data and associated attributes, which are easy to visualize, and it has been probed both by observations and simulations in the astrophysics literature. When a galaxy enters a high-density environment such as a group or cluster, its specific star formation rate (sSFR) is likely to drop. This process is known as environment quenching and represents a subset of the overall quenching process in galaxies \citep{2001AJ....122.1861S, 2003ApJS..149..289B, 2007ApJ...665..265F, 2007ApJS..173..342M, 2014MNRAS.440..889S} . This quenching is associated with a number of structural changes: chiefly, the increasing prominence of a central bulge. These effects have been well studied in the observations \citep{1972ApJ...176....1G, 1974ApJ...194....1O, 1976ApJ...208...13D, 2003MNRAS.341...33K, 2009MNRAS.393.1324B, 2010ApJ...721..193P, 2013MNRAS.428.3306W} and explored using numerical simulations \citep{1996Natur.379..613M, 1998ApJ...495..139M, 2009ApJ...694..789T, 2012MNRAS.423.1277D}. We view the results from both observational studies and simulations as a baseline to test whether our data-driven approach can lead us to similar results. If yes, then our approach has comparable utility for exploring astrophysical phenomena.

\section{Experiment} 
Following the general outline proposed earlier, we take a population of 26,706 galaxies in the redshift range $0.02<z<0.05$ with stellar masses $\log M_{\rm{stellar}} > 10.0$ from the Sloan Digital Sky Survey \citep{2000AJ....120.1579Y, 2003MNRAS.341...33K, 2004MNRAS.351.1151B, 2015ApJS..219...12A}. We train the Fader network on the environment by using samples of galaxies in the field and satellites in groups and clusters (exact criteria Table \ref{tab:criteria}) \citep{2007ApJ...671..153Y}. The underlying assumption is that field galaxies turn into satellite galaxies. We then take some field galaxies and use the trained network to show us what they would look like if they became satellites (Figure \ref{fig:hypothesis}).

\begin{figure}
\centering
\includegraphics[width=0.24\linewidth]{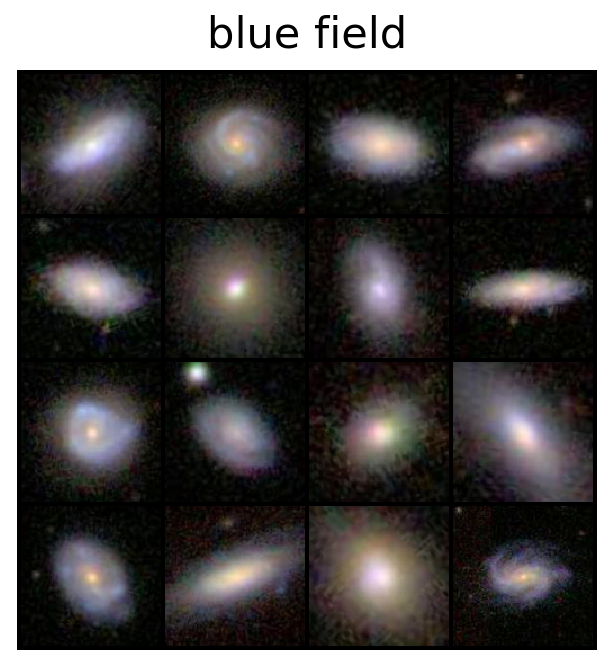}
\includegraphics[width=0.24\linewidth]{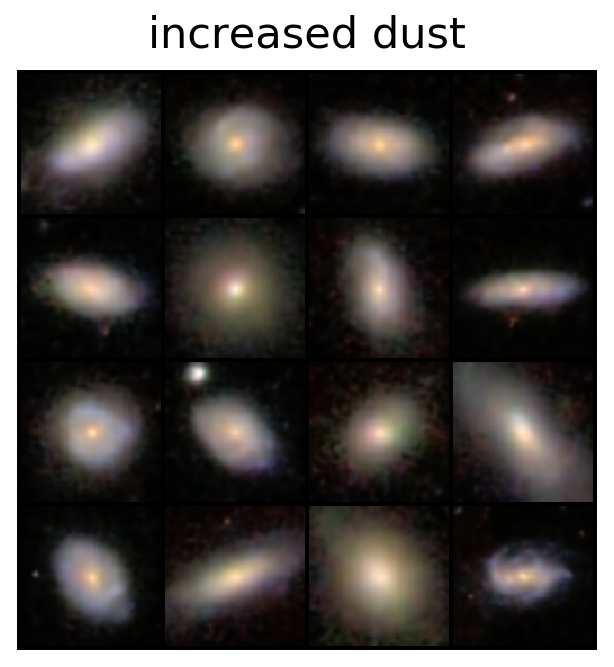}
\includegraphics[width=0.24\linewidth]{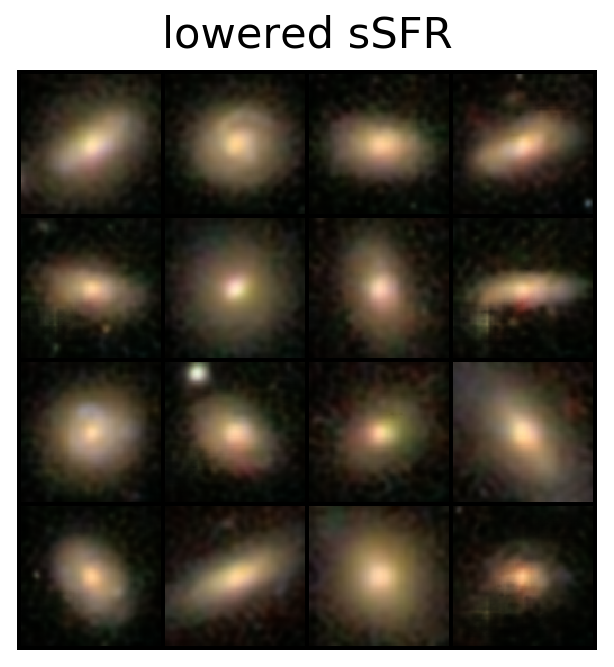}
\includegraphics[width=0.24\linewidth]{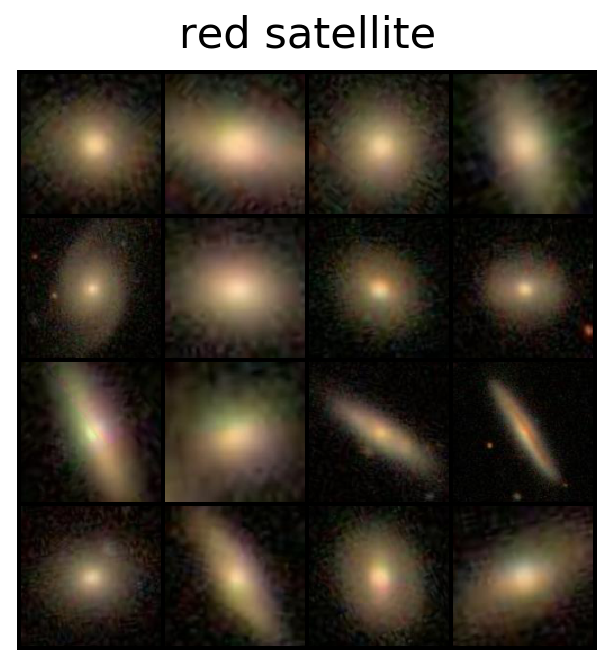}
\includegraphics[width=0.7\linewidth]{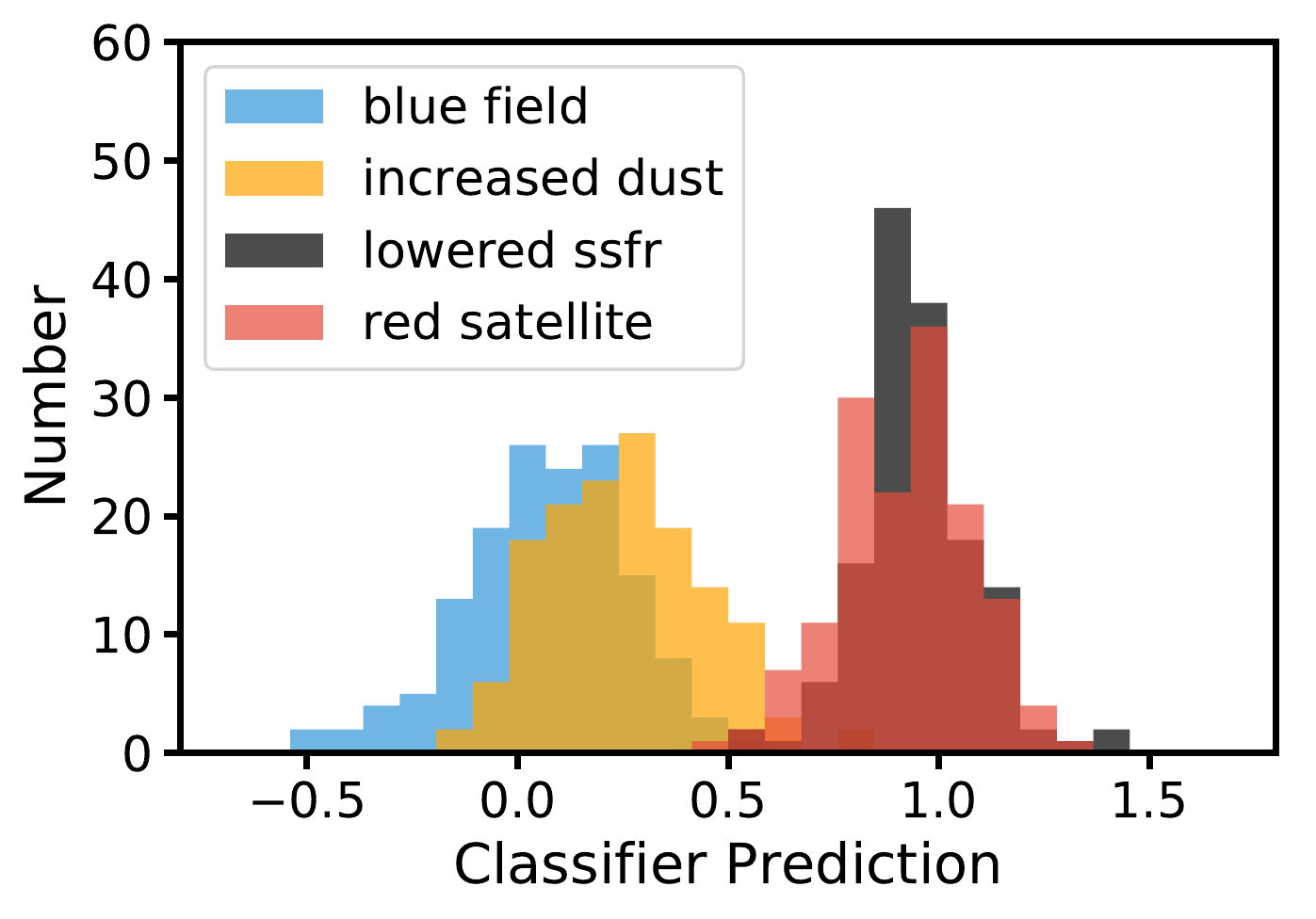}
\caption{\textit{(top)}  The original blue field galaxy images (left), and the real red satellites (right). We use the two Fader networks to increase the dust (left centre) and lower the sSFR (right centre). The galaxies with lowered sSFRs are qualitatively closer to the real red satellites. (\textit{bottom}) The same information from the top panel shown quantitatively using a ridge classifier. We show the classifier prediction distributions of the real blue field and red satellite distributions (blue and red, respectively), and the results of the two Fader transforms of the real blue field galaxies. The increased dust population (yellow) is very different from the real red satellites, while the lowered sSFR Fader population (grey) is very close. This suggests that lowering the sSFR is a better explanation for the formation of red satellite galaxies. }
\label{fig:results}
\end{figure}

Looking at Figure \ref{fig:hypothesis}, we note that as field galaxies become satellite galaxies, they become redder, and their bulges become more prominent. Using our domain knowledge, we can hypothesize two possible parameters which can explain the change in colour:
\begin{enumerate}
\item Shut down of SFR (quenching), or 
\item increased amount of dust.\end{enumerate}

To test these hypotheses, we train the Fader network with sSFR measurements from \cite{2003MNRAS.341...33K} using a sample of 11,240 SDSS galaxies, and with dust measurements from \cite{2011ApJS..195...13O} using a sample of 1,452 SDSS galaxies in the same redshift range (exact criteria Table \ref{tab:criteria}).

\begin{table}
\centering
\small
\begin{tabular}{ |l|l|l| }
\hline
 & label 0 & label 1 \\ \hline
\multirow{2}{*}{Environment} & $\log M_{halo}<11.8$ & $\log M_{halo}>12.5$ \\
 & $R_{proj_L} =0.0 \text{ kpc}$ & $1.0 \text{ kpc} <R_{proj_L}<500 \text{ kpc}$\\ \hline
sSFR & $\log sSFR>-10.5$ & $\log sSFR<-11.7$ \\ \hline
Dust & $ebvm_{gas}<0.15$ & $ebvm_{gas}>0.35$ \\ 
\hline
\end{tabular}
\caption{Overview of the selection criteria used to learn the environment, sSFR, and dust transform. Label 0 refers to field galaxies and label 1 refers to satellite galaxies.}
\label{tab:criteria}
\end{table}

That way we have learned two transforms, one which shuts down the SFR of a galaxy, and another one which increases the amount of dust. We verify that both sSFR and dust transform are learned correctly by training a ridge classifier  to predict the sSFR and dust physical property. The ridge classifier is a standard and widely used linear classifier. We train it on labelled images so that it predicts the correct label given the image. We adjust the regularization strength to maximise the accuracy of the classifier.  We see that both the sSFR and the dust distribution become shifted to the right, that is, lowered sSFR and increased dust.

We can now test these hypotheses by selecting a sample of 1,476 blue field galaxies ($u-r<1.9$) and 1,476 red satellite galaxies ($u-r>2.6$). We then train a ridge classifier to predict whether a galaxy is a blue field or a red satellite. We apply both transformations to a test set of 148 blue field galaxies, where we select the label parameter $y \in [0,1]$ such that the median of the transformed distribution lies closest to the median of the red satellite distribution. This results in $y_{sSFR}^*=0.64$ and $y_{dust}^*=1.00$.

We find that the classifier has difficulty in differentiating between the artificially lowered sSFR sample and the set of red satellites. The dust transform on the other hand is clearly not able to explain the colour transformation (Figure \ref{fig:results}). This supports the hypothesis that it is a change in sSFR which changes the colours and morphologies of galaxies as they enter high-density environments, in concordance with what we know from both observations and simulations.

What we have shown is that an astrophysicist can address a complex problem such as galaxy quenching, and make rapid progress in testing hypotheses using our method. Not only is training a neural network much faster and computationally less expensive than running a hydrodynamical simulation, it also does not rely on strong assumptions about the underlying physics, or suffer from limitations arising from coarse resolution. Similarly, our approach takes much greater advantage of the data than conventional model fitting to observations. Nevertheless, we were able to reach similarly robust conclusions to the baseline observational and simulation-based studies of the subject. 

\section{Limitations \& outlook}
Our approach also has some limitations. First, we `solved' an already fairly well-understood problem. We chose this because if we had used it to approach an unsolved problem and claimed new physical insight, it might not have been clear that either our method or the insight offered was reliable. This does not mean that our approach is necessarily limited to such well-understood problems. Second, we stress that our approach can only help us test hypotheses, not prove them in a mathematical sense. Third, as with any scientific observations, there is scope for confusion between real physical effects and deficiencies and biases in the training data and the network architecture. Finally, we highlight that this approach is not fully automatic, and domain knowledge by the user is still required.

Nevertheless, we believe our approach of using generative models like the Fader network to forward model physical processes and test hypotheses in a data-driven way has significant potential in astrophysics and other fields. Its central advantage is its data-driven nature which makes no assumptions on the underlying physics. As we have shown, human insight is still required for high-level interpretation.

\begin{acknowledgements}

K.S. acknowledges support from Swiss National Science Foundation Grants PP00P2\_138979 and PP00P2\_166159 and the ETH Zurich Department of Physics. C.Z. and the DS3Lab gratefully acknowledge the support from the Swiss National Science Foundation NRP 75 407540\_167266, IBM Zurich, Mercedes-Benz Research \& Development North America, Oracle Labs, Swisscom, Zurich Insurance, Chinese Scholarship Council, the Department of Computer Science at ETH Zurich, and the cloud computation resources from Microsoft Azure for Research award program. The SDSS data used, the Jupyter notebooks with the Fader network implementation, and instructions for how to run them can be found at \texttt{http://space.ml/proj/explore/}. 
\end{acknowledgements}

%
%

\bibliographystyle{aa}

\end{document}